\documentclass[twocolumn,showpacs,superscriptaddress,amsmath,amssymb,prb]{revtex4-1}

\usepackage[dvips]{graphicx}
\usepackage{dcolumn}
\usepackage{bm}
\usepackage{color}

\begin{document}

\title{Multiferroicity in the generic easy-plane triangular lattice antiferromagnet RbFe(MoO$_{4}$)$_{2}$}

\author{J.\,S.\,White,$^{1,2}$ Ch.\,Niedermayer,$^{1}$ G.\,Gasparovic,$^{3,4}$ C.\,Broholm,$^{3,4}$ J.\,M.\,S.\,Park,$^{5}$  A.\,Ya.\,Shapiro,$^{6}$ L.\,A.\,Demianets,$^{6}$ and M.\,Kenzelmann$^{7}$}

\affiliation{
Laboratory for Neutron Scattering, Paul Scherrer Institut, CH-5232 Villigen, Switzerland\\
$^{2}$ Laboratory for Quantum Magnetism, Ecole Polytechnique F\'{e}d\'{e}rale de Lausanne, CH-1015 Lausanne, Switzerland\\
$^{3}$ Institute for Quantum Matter and Department of Physics and Astronomy, The Johns Hopkins University, Baltimore, Maryland 21218, USA\\
$^{4}$ NIST Center for Neutron Research, National Institute of Standards and Technology, Gaithersburg, Maryland 20899, USA\\
$^{5}$ Neutron Science Division, Korea Atomic Energy Research Institute, Daejeon 305-353, Republic of Korea\\
$^{6}$ A. V. Shubnikov Institute for Crystallography RAS, 117333 Moscow, Russia\\
$^{7}$ Laboratory for Developments and Methods, Paul Scherrer Institut, CH-5232 Villigen, Switzerland}
\date{\today}

\begin{abstract}
RbFe(MoO$_{4}$)$_{2}$ is a quasi-two-dimensional (quasi-2D) triangular lattice antiferromagnet (TLA) that displays a zero-field magnetically-driven multiferroic phase with a chiral spin structure. By inelastic neutron scattering, we determine quantitatively the spin Hamiltonian. We show that the easy-plane anisotropy is nearly 1/3 of the dominant spin exchange, making RbFe(MoO$_{4}$)$_{2}$ an excellent system for studying the physics of the model 2D easy-plane TLA. Our measurements demonstrate magnetic-field induced fluctuations in this material to stabilize the generic finite-field phases of the 2D XY TLA. We further explain how Dzyaloshinskii-Moriya interactions can generate ferroelectricity \emph{only} in the zero field phase. Our conclusion is that multiferroicity in RbFe(MoO$_{4}$)$_{2}$, and its absence at high fields, results from the generic properties of the 2D XY TLA.
\end{abstract}





\maketitle
The two-dimensional (2D) triangular lattice antiferromagnet (TLA) is a prototypical model in which to study frustrated magnetic interactions. For easy-plane magnetic anisotropy, the 120$^{\circ}$ structure forms the zero-field groundstate. How the system evolves under an in-plane magnetic field has long been a subject of investigation since the equilibrium spin structures are expected to depend sensitively on both thermal~\citep{Kaw84,Kaw85,Kor86} and quantum fluctuations~\citep{Chu91}. Consequently, similar magnetic structure phase diagrams are expected within both XY and Heisenberg models~\citep{Kaw84,Kaw85,Kor86,Lee86,Chu91,Bou96,Sea11a,Sea11b,Gvo11}. Furthermore, a number of TLAs have multiferroic ground states, but the role of the triangular magnetic topology in the emergence of ferroelectricity is not well understood.

While experimental realizations with which to test the predictions of the 2D TLA models are rare, RbFe(MoO$_{4}$)$_{2}$  (RFMO) stands out as an excellent example of a quasi-2D easy-plane TLA~\citep{Jor04,Svi03,Svi06,Smi07,Ken07}. In addition, the zero field magnetically-ordered phase of RFMO is ferroelectric~\citep{Ken07,Hea12}, so the material provides a unique opportunity to study how multiferroicity is related to the generic fluctuations of the easy-plane TLA.

\begin{figure}
\includegraphics[width=0.48\textwidth]{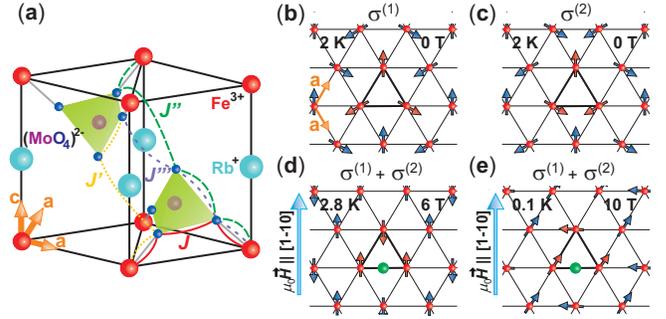}
\caption{(Color online) (a) The low temperature $P\overline{3}$ structure of RbFe(MoO$_{4}$)$_{2}$. O$^{2-}$ mediated superexchange interaction paths are indicated, and the exchange hierarchy is $J\gg J'>J''\sim J'''$. (b)-(e) Magnetic structures within a single triangular lattice layer for fields up to 10~T~\citep{Ken07}. (b) and (c) show degenerate zero field 120$^{\circ}$ spin structures of anti-phase chirality. Each can be described by a single phenomenological order parameter (b) $\sigma^{(1)}$ or (c) $\sigma^{(2)}$~\citep{Ken07,Har07}. The magnetic structures under $\mu_{0}\vec{H}$$\parallel$[1-10] of (d) 6~T and (e) 10~T are each described by a combination of $\sigma^{(1)}$ and $\sigma^{(2)}$. Green circles show inversion centers, and red spin triangles are representative for each magnetic structure.}
\label{fig:structures}
\end{figure}

In RFMO, magnetic Fe$^{3+}$ ions ($S$=5/2) form equilateral triangular lattice planes stacked along the \textbf{c}-axis [Fig.~\ref{fig:structures}~(a)]. In zero field, for $T<T_{\rm N}\sim$~3.8~K the system displays incommensurate (IC) proper screw order with a 120$^{\circ}$ structure in plane. The corresponding wave vector \textbf{Q}~=~(1/3,1/3,$q_{z}$), where $q_{z}\sim0.46$~\citep{Ken07}. The 120$^{\circ}$ structure is $\emph{chiral}$, since for any spin triangle there are two equivalent, yet distinct, ways to arrange the spins [see Figs.~\ref{fig:structures}~(b) and (c)]. These chiral 120$^{\circ}$ structures break the crystal inversion symmetry $\mathcal {I}$, and generate a spontaneous ferroelectric polarization along the \textbf{c}-axis, $P_{\rm c}$, the direction of which ($\pm P_{\rm c}$) depends on the sense of chirality~\citep{Ken07,Hea12}. Here we parameterize spin chirality locally for any spin triangle by $\mathcal{K}=(2/3\sqrt{3})(\textbf{S}_{1}\times \textbf{S}_{2}+\textbf{S}_{2}\times \textbf{S}_{3}+\textbf{S}_{3}\times \textbf{S}_{1})\cdot \hat{\textbf{z}}/S^2$. With this definition, $\mathcal{K}=\pm1$ for spin triangles of the 120$^{\circ}$ structure. Recently, it was identified that the crystal distortion which sets in for $T^{\ast}\lesssim 190$~K~\citep{Ina07} is `ferroaxial'~\citep{Joh11,Hea12}, and enables a symmetric-exchange coupling between the magnetic helicity (sign of $q_{z}$) and the sense of triangular chirality~\cite{Hea12}. While the sign of the crystal distortion in each ferroaxial domain fixes the possible relationships between magnetic helicity and triangular chirality~\citep{Har07,Hea12}, the direction of $P_{\rm c}$ is always determined by the sense of chirality. Figs.~\ref{fig:structures}~(b) and (c) show single 120$^{\circ}$ spin structure planes of anti-phase chirality that will generate a $P$ along the opposing directions $\pm P_{\rm c}$~\citep{Ken07,Hea12}.


For easy-plane magnetic fields $\mu_{0}\vec{H}$$\parallel$[1-10], the chiral-ordered multiferroic phase is replaced by a paraelectric (PE) and commensurate (C) phase with \textbf{Q}~=~(1/3,1/3,1/3). The magnetic structure at $\mu_{0}H$=6~T is collinear [Fig.~\ref{fig:structures}(d)] with two spins on each spin triangle parallel to $\mu_{0}\vec{H}$, and the remaining spin antiparallel. This structure is expected for the classical TLA in applied fields close to 1/3$H_{\rm s}$, where $H_{\rm s}$ is the saturation field~\citep{Kaw85,Kor86,Lee86,Chu91}. Close to the upper field limit of the C phase at $\mu_{0}H$=10~T, the refined structure [Fig.~\ref{fig:structures}(e)] is the expected `two-up one-down' arrangement, with just two parallel spins on each spin triangle. Since all C magnetic structures display two parallel spins on each spin triangle, $\mathcal{K}=0$ everywhere. Above 10~T, the C phase is replaced by a high field incommensurate (HFI) phase with \textbf{Q}~=~(1/3,1/3,$q_{z}$) the microscopic properties of which are not yet reported. Since such a high field phase is unexpected theoretically, open questions persist regarding both the origin of the phase transition, and the relation between the magnetic and electric properties.

Here we report neutron scattering studies of the microscopic magnetism in RFMO. The spin wave dispersion is measured and used to extract a spin Hamiltonian that quantifies both the magnetic interactions and a large XY anisotropy. Elastic measurements reveal field-induced fluctuations to cause both the high field transitions, and the suppression of ferroelectricity. Our refinement of the HFI magnetic structure shows that it is not chiral, which is consistent with a bulk paraelectric state. We also show that the high field C-IC transition is not a generic property of the XY TLA. Finally, we discuss the origin of multiferroicity in this system.

\begin{figure}
\includegraphics[width=0.48\textwidth]{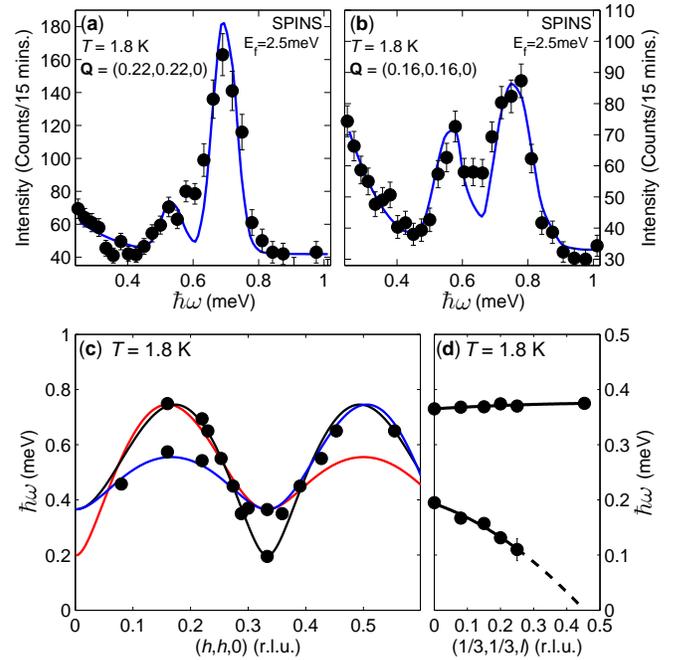}
\caption{(Color online) Constant wavevector scans of the spin wave excitations at zero field, and $T=1.8$~K at (a) \textbf{Q}~=~($0.22$,$0.22$,$0$), and (b) \textbf{Q}~=~($0.16$,$0.16$,$0$). The fit lines result from a numerical convolution of Lorentzian energy profiles with the spectrometer resolution function. (c) The in-plane dispersion constructed from the analysis of the constant ($h$,$h$,$0$) wavevector scans. The lines in (c) show the dispersion of the three modes described by the spin Hamiltonian given in the text. (d) The inter-plane dispersion determined along (1/3,1/3,$l$) with lines as guides for the eye.}
\label{fig:spin_waves}
\end{figure}

Inelastic neutron scattering measurements were performed using the SPINS instrument at NIST, USA. Single crystals of RFMO were synthesized using a flux method~\citep{Kle70}, and a mosaic with mass 80~mg was coaligned and mounted with a ($h$,$h$,$l$) horizontal scattering plane. Figs.~\ref{fig:spin_waves}~(a)-(b) show typical energy scans measured at $T=1.8$~K in the multiferroic phase, and with constant ($h$,$h$,0) wavevector. To extract the spin-wave mode energies from the scans, we numerically convoluted Lorentzian energy profiles with the spectrometer resolution function. The computed lines shown in Figs.~\ref{fig:spin_waves}~(a)-(b) reveal each scan to display two spin wave modes. By determining the peak positions in all energy scans, the in-plane dispersion relation along ($h$,$h$,0) was obtained [Fig.~\ref{fig:spin_waves}(c)]. Similar scans also allowed the determination of the inter-plane dispersion along (1/3,1/3,$l$) [Fig.~\ref{fig:spin_waves}(d)]. Here, two weakly dispersive modes indicate the 2D nature of the system. The dispersion of the low energy mode evidences interplane interactions that stabilize 3D magnetic order. Furthermore, its wavevector dependence is consistent with a Goldstone mode emerging from the magnetic Bragg wavevector (1/3,1/3,$q_{z}$) as expected for a magnetic state that breaks the in-plane rotational symmetry.

Using linear spin-wave theory, the in-plane dispersion was calculated using the following Hamiltonian relation
\begin{equation}
\mathcal{H}=J\sum_{\langle i,j\rangle}\textbf{S}_{i}\cdot \textbf{S}_{j}+D\sum_{i}S_{i}^{z}S_{i}^{z}+J_{\rm p}\sum_{\langle i,k\rangle}\textbf{S}_{i}\cdot \textbf{S}_{k}.
\label{Hamiltonian}
\end{equation}
Here, $J$ is the nearest-neighbor antiferromagnetic Heisenberg exchange, the sum is over all in-plane nearest-neighbor pairs, and $D$ is the single ion anisotropy. Since there are at least three inter-plane interactions that can not all be determined individually from our experiments, we approximate these by an `effective' nearest-neighbor inter-plane interaction $J_{\rm p}$~\citep{Sup2}. Following the approach of Refs.~\onlinecite{Vaj05,Har92}, after a standard diagonalization of the linearized form of $\mathcal{H}$, three modes are expected in the spin-wave dispersion~\citep{Sup}. In Fig.~\ref{fig:spin_waves}~(c) we show that this is consistent with the data for RFMO, and that the simple model describes the in-plane dispersion extremely well, with $J=0.086(2)$~meV, $D=0.027(1)$~meV and $J_{\rm p}=0.0007(1)$~meV. These results establish that RFMO is an XY-like ($D/J=0.31(1)$) and 2D ($J_{\rm p}/J=0.008(1)$) TLA. The relatively large value for $D$ in the proposed Hamiltonian ensures that the magnetic moments remain in the plane, avoiding a spin-flop transition to a state where magnetic moments perpendicular to $\mu_{0}\vec{H}$ point along the \textbf{c}-axis. This is a crucial property of the XY TLA under high in-plane fields.

\begin{figure}
\includegraphics[width=0.48\textwidth]{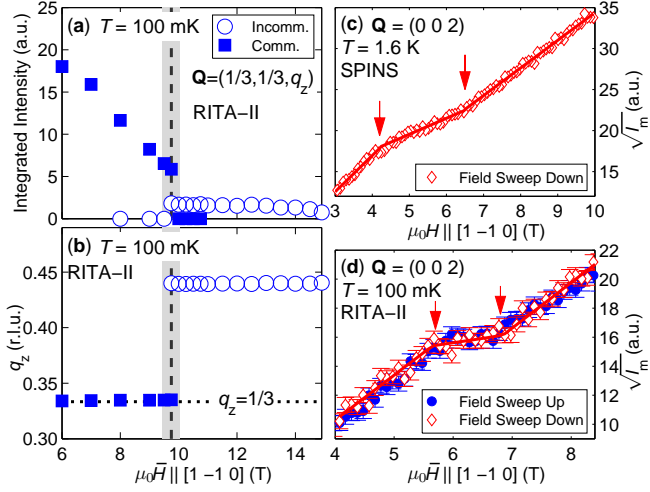}
\caption{(Color online) The $\mu_{0}H$-dependence at $T=100$~mK of (a) the neutron integrated intensity recorded at the Q=(1/3,1/3,$q_{\rm z}$) position, and (b) the $q_{z}$ component. Dashed lines mark the transition fields between different phases, with the uncertainties indicated by the shaded regions. At (c) $T=1.6$~K and (d) $T=100$~mK we show the $\mu_{0}H$-dependence of the square-root of the magnetic neutron intensity, $\sqrt{I_{\rm m}}$ measured at the (002) position. Red arrows indicate the field range of the intensity plateaus.}
\label{fig:H_dep_data}
\end{figure}

Elastic neutron diffraction experiments were performed using the RITA-II instrument, at PSI, Switzerland. A single 6.5~mg crystal from the inelastic measurements was mounted with an ($h$,$h$,$l$) horizontal scattering plane, and installed inside a 14.9~T vertical field cryomagnet with dilution refrigerator. Measurements of the $\mu_{0}H$- and $T$-dependence of the magnetic order are consistent with previous work~\citep{Ken07}. Figs.~\ref{fig:H_dep_data}(a) and (b) show the $\mu_{0}H$-dependence of magnetic order in the previously unexplored portion of the phase diagram for 6~T$<\mu_{0}H<$14.9~T at $T$=100~mK. A discontinuous transition clearly separates the intermediate field C and HFI phases. For the latter phase, the values of $q_{z}$ are similar to those reported previously at 2.8~K~\citep{Ken07}, and no significant $\mu_{0}H$-dependence is observed.

\begin{figure}
\includegraphics[width=0.48\textwidth]{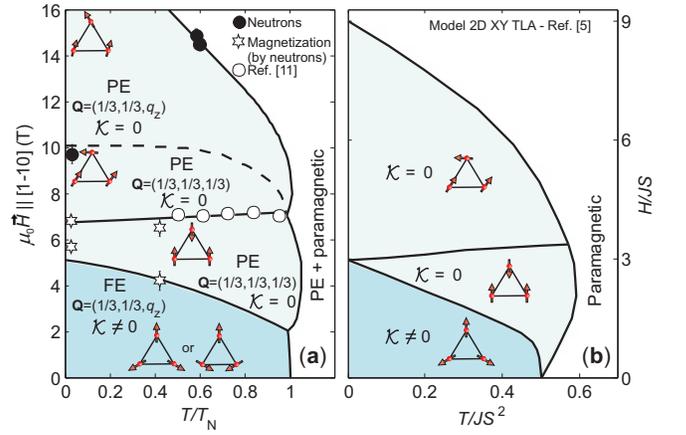}
\caption{(Color online) (a) An updated schematic of the $\mu_{0}\vec{H}$$\parallel$[1-10] versus reduced temperature ($T/T_{\rm N}$) phase diagram of RFMO first shown in Ref.~\citep{Ken07}. Open circles show the end of magnetization plateaus determined in Ref.~\citep{Svi03}, and stars show the start and end of similar plateaus determined using neutrons [Figs.~\ref{fig:H_dep_data}(c) and (d)]. Ferroelectric (FE) and paraelectric (PE) phases are indicated. The solid (dashed) lines indicate magnetic phase boundaries consistent (inconsistent) with those calculated for a model 2D XY TLA shown in panel (b)~\citep{Lee86}. Phase diagrams similar to that shown in (b) were obtained from other XY~\citep{Kor86,Chu91,Bou96} and Heisenberg~\citep{Kaw84,Kaw85,Chu91,Sea11b,Gvo11} model calculations.}
\label{fig:phase_diagram}
\end{figure}

The fluctuations for easy-plane magnetic fields are further characterized by measurements of the $\mu_{0}H$-dependent magnetic neutron intensity, $I_{\rm m}$ at the (002) nuclear position. At both $T=1.6$~K and $T=100$~mK [Figs.~\ref{fig:H_dep_data}(c) and (d)] we show the $\mu_{0}H$-dependence of $\sqrt{I_{\rm m}}$, since this quantity provides a direct measure of the field-induced bulk magnetization. At both temperatures $\sqrt{I_{\rm m}}$ depends linearly on $\mu_{0}H$ over most of the field range. At intermediate fields however, intensity plateaus are observed that correspond to the hallmark 1/3 magnetization plateaus expected when the collinear structure is stabilized close to 1/3$H_{\rm s}$~\citep{Kor86,Chu91,Svi03}, where $H_{\rm s}\sim$~19~T~\citep{Svi03,Svi06,Smi07}. In particular, the plateau at $T$~=~100~mK, which is reported here for the first time at such low temperature, occupies the finite field range 5.7-6.7 T. This is much narrower than similar plateaus measured at higher temperature, such as that shown in Fig.~\ref{fig:H_dep_data}(c), or by SQUID magnetometry~\citep{Svi03,Svi06,Smi07}. These observations confirm the expectation that increased thermal fluctuations stabilize the collinear structure [Fig.~\ref{fig:structures}(d)] over a wider field range~\citep{Kor86,Chu91}. Using the data shown in Figs.~\ref{fig:H_dep_data}(a)-(d), in Fig.~\ref{fig:phase_diagram}(a) we present an updated version of the phase diagram first presented in Ref.~\citep{Ken07}.

Next we discuss the refinement of the magnetic structure within the HFI phase. The results are analyzed within the phenomenological framework developed in Refs.~\onlinecite{Ken07,Har07}, and we note that a similar approach was recently proposed~\citep{Rib11}. In RFMO, all easy-plane spin structures can be described in terms of two complex-valued scalar order parameters $\sigma^{(1)}$ and $\sigma^{(2)}$, that correspond to amplitudes for 120$^{\circ}$ spin structures of opposite chirality as depicted in Figs.~\ref{fig:structures}(a) and (b). The order parameters enter directly into the part of the free energy $\mathcal{F}$, that successfully describes the magnetoelectric coupling;
\begin{equation}
\mathcal{F}\propto K\left[|\sigma^{(1)}\left(q_{z}\right)|^{2}-|\sigma^{(2)}\left(q_{z}\right)|^{2}\right]P_{c}.
\label{Trilinear}
\end{equation}
Here $K$ is a symmetry-independent coupling constant. A 120$^{\circ}$ spin structure is described by either $\sigma^{(1)}\left(q_{z}\right)\neq0$ and $\sigma^{(2)}\left(q_{z}\right)=0$, or $\sigma^{(1)}\left(q_{z}\right)=0$ and $\sigma^{(2)}\left(q_{z}\right)\neq0$. In these cases, the observed $P_{c}$ is expected in accord with Eq.~\ref{Trilinear}. In contrast, the C magnetic structures are each described by $|\sigma^{(1)}|^{2}=|\sigma^{(2)}|^{2}$, which according to Eq.~\ref{Trilinear} is consistent with a bulk PE state.

\begin{figure}
\includegraphics[width=0.48\textwidth]{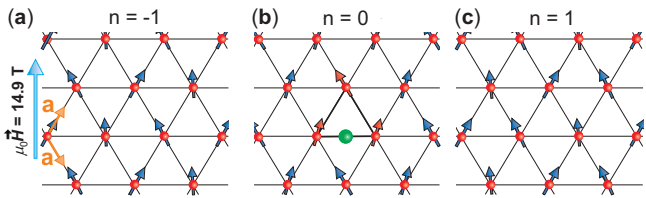}
\caption{(Color online) Panels (a)-(c) show the refined incommensurate magnetic structure at $\mu_{0}H\parallel$[1-10]~=~14.9~T and $T$~=~100~mK for adjacent layers along the \textbf{c}-axis, when $\sigma^{(1)}$=-$\sigma^{(2)}=0.94$. In (b), the red spin triangle is representative for the magnetic structure, and the green circle indicates an inversion center.}
\label{fig:structure_refinement}
\end{figure}

The magnetic structure in the HFI phase was determined at $\mu_{0}H$~=~14.9~T and $T$~=~100~mK. The integrated intensities of 36 magnetic peaks were collected and found to be best described by $\sigma^{(1)}$~=~0.94(4) and $\sigma^{(2)}$=-0.94(4)-$i$0.00(20) with $\chi^{2}=2.97$ and $R=0.33$~\citep{Sup}. Since within uncertainty $|\sigma^{(1)}|^{2}=|\sigma^{(2)}|^{2}$, a $P_{c}$ is not expected according to Eq.~\ref{Trilinear}. The refined magnetic structure is shown in Figs.~\ref{fig:structure_refinement}(a)-(c) for adjacent layers along the \textbf{c}-axis. Unlike all lower field phases, the moments are weakly amplitude modulated and, while they tend to order collinearly along $[110]$ in the plane orthogonal to $\mu_{0}H$, they are strongly canted along the field direction. The moment magnitude determined from the refinement is 4.0(5)+0.2(3)sin($q_{z}n\textbf{c}$+$\phi$)$\mu_{\rm B}$ where the integer $n$ indexes spin planes displaced along \textbf{c}, and the modulation period is $\sim$~65~\AA~($\sim$9\textbf{c}). Importantly, the refined HFI magnetic structure both preserves $\mathcal{I}$ and displays $\mathcal{K}=0$ for every spin triangle. Indeed for any spin of the refined structure, the modulation phases of the other two spins on the spin triangle exhibit relative values $\phi$, of $+2\pi/3$ and $-2\pi/3$. These phase differences perfectly preserve $\mathcal{K}=0$ everywhere since, unlike the unmodulated C structures, this condition can be satisfied without requiring at least two parallel spins [Figs.~\ref{fig:structure_refinement}(a) and (c)].

Using Fig.~\ref{fig:phase_diagram} we compare between the experimental phase diagram for RFMO (panel (a)) and the expected phase diagram for a model XY TLA (panel (b))~\citep{Lee86}. For purely classical spins, in both XY~\citep{Lee86,Chu91} and Heisenberg~\citep{Kaw84,Kaw85} models the magnetization plateau is expected to collapse to the singular field 1/3~$H_{\rm s}$ in the $T\rightarrow$~0 limit~\citep{Kaw85,Kor86,Lee86} [see Fig.~\ref{fig:phase_diagram}(b)]. As indicated in Fig.~\ref{fig:phase_diagram}(a), our observation of a finite $\sim$~1~T plateau at T~=~100~mK strongly suggests that the plateau occupies a finite field range as $T\rightarrow$~0. These measurements could indicate quantum fluctuations~\citep{Chu91} to take the place of thermal fluctuations in stabilizing the 1/3 magnetization plateau. However, since RFMO is an $S=5/2$, strictly quasi-2D system, biquadratic interactions likely play an important role in stabilizing the plateau~\citep{Pen04}. Nonetheless, by combining our observations across the phase diagram with the description of the spin dynamics that evidence the large XY anisotropy, we confirm the properties of quasi-2D RFMO to display remarkable agreement with the predictions of the model 2D XY TLA.

To unravel the relationship between multiferroicity in RFMO and the generic properties of the 2D XY TLA, we see from Fig.~\ref{fig:phase_diagram}(a) that the low field IC-C transition essentially separates the multiferroic chiral and PE collinear phases (notwithstanding subtle phase modifications within the vicinity of the transition~\citep{Svi03}). Therefore, the field-driven collapse of ferroelectricity occurs as a consequence of the field-induced fluctuations expected for the generic 2D XY TLA, and is understood in terms of a suppressed chiral symmetry~\citep{Kaw85,Chu91}. In the high field half of the phase diagram, the in-plane physics and symmetry remain dominated by properties expected for the model 2D XY TLA. This is evidenced by the observation that all magnetic structures in the C and HFI phases preserve $\mathcal{I}$ and display the expected property that $\mathcal{K}=0$ everywhere~\citep{Kaw85,Kor86,Lee86,Chu91}. It therefore follows that the high-field C-IC transition observed in RFMO is not a generic property of the 2D XY TLA, but instead occurs as a consequence of weaker interactions in the full quasi-2D Hamiltonian.

Finally, we discuss the multiferroic mechanism in RFMO. Symmetry-based phenomenological approaches successfully explain the emergence of $P_{c}$~\citep{Ken07,Har07,Rib11,Hea12}, yet the definitive microscopic origin remains to be clarified. Here, we consider the role of the Dzyaloshinskii-Moriya interaction (DMI) which can exist between non-collinear magnetic moments on nearest in-plane neighbors that locally break $\mathcal{I}$. According to the well-known inverse DMI/spin current model $\textbf{\textrm{P}}_{ij}\propto\textbf{\textrm{e}}_{ij}\times(\textbf{\textrm{S}}_{i} \times \textbf{\textrm{S}}_{j})$ \citep{Ser06,Kat05}, the $P$ is expected in the triangular lattice plane, and so can not explain the origin of $P_{\rm c}$. However, since in RFMO the unit vector between nearest in-plane neighbors $i$ and $j$ neither includes a mirror plane, nor is $\perp$ to a two-fold rotation axis, an additional polarization component $\perp$ to the triangular lattice plane, $p_{\rm c}\propto(\textbf{\textrm{S}}_{i} \times \textbf{\textrm{S}}_{j})$ is allowed by symmetry~\citep{Kap11}. By evaluating the products between all nearest in-plane neighbours, we find that the sum of $p_{\rm c}$ terms to be finite \emph{only} for the chiral 120$^{\circ}$ structure, and otherwise vanish for every 2D \emph{plane} of the experimentally observed magnetic structures at higher field. This strongly suggests that the DMI between in-plane neighbors is crucial for the emergence of ferroelectricity.

In summary, our neutron scattering experiments establish the quasi-2D triangular lattice antiferromagnet (TLA) RbFe(MoO$_{4}$)$_{2}$ to display many properties in remarkable agreement with those of the model 2D XY TLA. We demonstrate the magnetic phase diagram to be dominated by field-induced fluctuations, and that the bulk multiferroic state arises as a consequence of the generic properties of the model 2D XY TLA. We further identify a possible Dzyaloshinskii-Moriya interaction in the chiral-ordered multiferroic phase that may give rise to multiferroicity in this system.

Experiments were performed at the Swiss spallation neutron source SINQ, Paul Scherrer Institut, Switzerland. Financial support from the Swiss NCCR program MaNEP is gratefully acknowledged. Work at IQM was supported by the US Department of Energy, Office of Basic Energy Sciences, Division of Materials Sciences and Engineering, under award DE-FG02-08ER46544.

\bibliography{rfmo}

\begin{thebibliography}{28}%
\makeatletter
\providecommand \@ifxundefined [1]{%
 \@ifx{#1\undefined}
}%
\providecommand \@ifnum [1]{%
 \ifnum #1\expandafter \@firstoftwo
 \else \expandafter \@secondoftwo
 \fi
}%
\providecommand \@ifx [1]{%
 \ifx #1\expandafter \@firstoftwo
 \else \expandafter \@secondoftwo
 \fi
}%
\providecommand \natexlab [1]{#1}%
\providecommand \enquote  [1]{``#1''}%
\providecommand \bibnamefont  [1]{#1}%
\providecommand \bibfnamefont [1]{#1}%
\providecommand \citenamefont [1]{#1}%
\providecommand \href@noop [0]{\@secondoftwo}%
\providecommand \href [0]{\begingroup \@sanitize@url \@href}%
\providecommand \@href[1]{\@@startlink{#1}\@@href}%
\providecommand \@@href[1]{\endgroup#1\@@endlink}%
\providecommand \@sanitize@url [0]{\catcode `\\12\catcode `\$12\catcode
  `\&12\catcode `\#12\catcode `\^12\catcode `\_12\catcode `\%12\relax}%
\providecommand \@@startlink[1]{}%
\providecommand \@@endlink[0]{}%
\providecommand \url  [0]{\begingroup\@sanitize@url \@url }%
\providecommand \@url [1]{\endgroup\@href {#1}{\urlprefix }}%
\providecommand \urlprefix  [0]{URL }%
\providecommand \Eprint [0]{\href }%
\providecommand \doibase [0]{http://dx.doi.org/}%
\providecommand \selectlanguage [0]{\@gobble}%
\providecommand \bibinfo  [0]{\@secondoftwo}%
\providecommand \bibfield  [0]{\@secondoftwo}%
\providecommand \translation [1]{[#1]}%
\providecommand \BibitemOpen [0]{}%
\providecommand \bibitemStop [0]{}%
\providecommand \bibitemNoStop [0]{.\EOS\space}%
\providecommand \EOS [0]{\spacefactor3000\relax}%
\providecommand \BibitemShut  [1]{\csname bibitem#1\endcsname}%
\let\auto@bib@innerbib\@empty
\bibitem [{\citenamefont {Kawamura}(1984)}]{Kaw84}%
  \BibitemOpen
  \bibfield  {author} {\bibinfo {author} {\bibfnamefont {H.}~\bibnamefont
  {Kawamura}},\ }\href@noop {} {\bibfield  {journal} {\bibinfo  {journal} {J.
  Phys. Soc. Jpn.}\ }\textbf {\bibinfo {volume} {53}},\ \bibinfo {pages} {2452}
  (\bibinfo {year} {1984})}\BibitemShut {NoStop}%
\bibitem [{\citenamefont {Kawamura}\ and\ \citenamefont
  {Miyashita}(1985)}]{Kaw85}%
  \BibitemOpen
  \bibfield  {author} {\bibinfo {author} {\bibfnamefont {H.}~\bibnamefont
  {Kawamura}}\ and\ \bibinfo {author} {\bibfnamefont {S.}~\bibnamefont
  {Miyashita}},\ }\href@noop {} {\bibfield  {journal} {\bibinfo  {journal} {J.
  Phys. Soc. Jpn.}\ }\textbf {\bibinfo {volume} {54}},\ \bibinfo {pages} {4530}
  (\bibinfo {year} {1985})}\BibitemShut {NoStop}%
\bibitem [{\citenamefont {Korshunov}(1986)}]{Kor86}%
  \BibitemOpen
  \bibfield  {author} {\bibinfo {author} {\bibfnamefont {S.~E.}\ \bibnamefont
  {Korshunov}},\ }\href@noop {} {\bibfield  {journal} {\bibinfo  {journal} {J.
  Phys. C: Solid. State Phys.}\ }\textbf {\bibinfo {volume} {19}},\ \bibinfo
  {pages} {5927} (\bibinfo {year} {1986})}\BibitemShut {NoStop}%
\bibitem [{\citenamefont {Chubukov}\ and\ \citenamefont
  {Golosov}(1991)}]{Chu91}%
  \BibitemOpen
  \bibfield  {author} {\bibinfo {author} {\bibfnamefont {A.~V.}\ \bibnamefont
  {Chubukov}}\ and\ \bibinfo {author} {\bibfnamefont {D.~I.}\ \bibnamefont
  {Golosov}},\ }\href@noop {} {\bibfield  {journal} {\bibinfo  {journal} {J.
  Phys.: Condens. Matter}\ }\textbf {\bibinfo {volume} {3}},\ \bibinfo {pages}
  {69} (\bibinfo {year} {1991})}\BibitemShut {NoStop}%
\bibitem [{\citenamefont {Lee}\ \emph {et~al.}(1986)\citenamefont {Lee},
  \citenamefont {Joannopoulos}, \citenamefont {Negele},\ and\ \citenamefont
  {Landau}}]{Lee86}%
  \BibitemOpen
  \bibfield  {author} {\bibinfo {author} {\bibfnamefont {D.~H.}\ \bibnamefont
  {Lee}}, \bibinfo {author} {\bibfnamefont {J.~D.}\ \bibnamefont
  {Joannopoulos}}, \bibinfo {author} {\bibfnamefont {J.~W.}\ \bibnamefont
  {Negele}}, \ and\ \bibinfo {author} {\bibfnamefont {D.~P.}\ \bibnamefont
  {Landau}},\ }\href@noop {} {\bibfield  {journal} {\bibinfo  {journal} {Phys.
  Rev. B}\ }\textbf {\bibinfo {volume} {33}},\ \bibinfo {pages} {450} (\bibinfo
  {year} {1986})}\BibitemShut {NoStop}%
\bibitem [{\citenamefont {Boubcheur}\ \emph {et~al.}(1996)\citenamefont
  {Boubcheur}, \citenamefont {Loison},\ and\ \citenamefont {Diep}}]{Bou96}%
  \BibitemOpen
  \bibfield  {author} {\bibinfo {author} {\bibfnamefont {E.~H.}\ \bibnamefont
  {Boubcheur}}, \bibinfo {author} {\bibfnamefont {D.}~\bibnamefont {Loison}}, \
  and\ \bibinfo {author} {\bibfnamefont {H.~T.}\ \bibnamefont {Diep}},\
  }\href@noop {} {\bibfield  {journal} {\bibinfo  {journal} {Phys. Rev. B}\
  }\textbf {\bibinfo {volume} {54}},\ \bibinfo {pages} {4165} (\bibinfo {year}
  {1996})}\BibitemShut {NoStop}%
\bibitem [{\citenamefont {Seabra}\ and\ \citenamefont
  {Shannon}(2011)}]{Sea11a}%
  \BibitemOpen
  \bibfield  {author} {\bibinfo {author} {\bibfnamefont {L.}~\bibnamefont
  {Seabra}}\ and\ \bibinfo {author} {\bibfnamefont {N.}~\bibnamefont
  {Shannon}},\ }\href@noop {} {\bibfield  {journal} {\bibinfo  {journal} {Phys.
  Rev. B}\ }\textbf {\bibinfo {volume} {83}},\ \bibinfo {pages} {134412}
  (\bibinfo {year} {2011})}\BibitemShut {NoStop}%
\bibitem [{\citenamefont {Seabra}\ \emph {et~al.}(2011)\citenamefont {Seabra},
  \citenamefont {Momoi}, \citenamefont {Sindzingre},\ and\ \citenamefont
  {Shannon}}]{Sea11b}%
  \BibitemOpen
  \bibfield  {author} {\bibinfo {author} {\bibfnamefont {L.}~\bibnamefont
  {Seabra}}, \bibinfo {author} {\bibfnamefont {T.}~\bibnamefont {Momoi}},
  \bibinfo {author} {\bibfnamefont {P.}~\bibnamefont {Sindzingre}}, \ and\
  \bibinfo {author} {\bibfnamefont {N.}~\bibnamefont {Shannon}},\ }\href@noop
  {} {\bibfield  {journal} {\bibinfo  {journal} {Phys. Rev. B}\ }\textbf
  {\bibinfo {volume} {84}},\ \bibinfo {pages} {214418} (\bibinfo {year}
  {2011})}\BibitemShut {NoStop}%
\bibitem [{\citenamefont {Gvozdikova}\ \emph {et~al.}(2011)\citenamefont
  {Gvozdikova}, \citenamefont {Melchy},\ and\ \citenamefont
  {Zhitomirsky}}]{Gvo11}%
  \BibitemOpen
  \bibfield  {author} {\bibinfo {author} {\bibfnamefont {M.~V.}\ \bibnamefont
  {Gvozdikova}}, \bibinfo {author} {\bibfnamefont {P.-E.}\ \bibnamefont
  {Melchy}}, \ and\ \bibinfo {author} {\bibfnamefont {M.~E.}\ \bibnamefont
  {Zhitomirsky}},\ }\href@noop {} {\bibfield  {journal} {\bibinfo  {journal}
  {J. Phys.: Condens. Matter}\ }\textbf {\bibinfo {volume} {23}},\ \bibinfo
  {pages} {164209} (\bibinfo {year} {2011})}\BibitemShut {NoStop}%
\bibitem [{\citenamefont {Jorge}\ \emph {et~al.}(2004)\citenamefont {Jorge},
  \citenamefont {Capan}, \citenamefont {Ronning}, \citenamefont {Jaime},
  \citenamefont {Kenzelmann}, \citenamefont {Gasparovic}, \citenamefont
  {Broholm}, \citenamefont {Shapiro},\ and\ \citenamefont {Demianets}}]{Jor04}%
  \BibitemOpen
  \bibfield  {author} {\bibinfo {author} {\bibfnamefont {G.~A.}\ \bibnamefont
  {Jorge}}, \bibinfo {author} {\bibfnamefont {C.}~\bibnamefont {Capan}},
  \bibinfo {author} {\bibfnamefont {F.}~\bibnamefont {Ronning}}, \bibinfo
  {author} {\bibfnamefont {M.}~\bibnamefont {Jaime}}, \bibinfo {author}
  {\bibfnamefont {M.}~\bibnamefont {Kenzelmann}}, \bibinfo {author}
  {\bibfnamefont {G.}~\bibnamefont {Gasparovic}}, \bibinfo {author}
  {\bibfnamefont {C.}~\bibnamefont {Broholm}}, \bibinfo {author} {\bibfnamefont
  {A.~Y.}\ \bibnamefont {Shapiro}}, \ and\ \bibinfo {author} {\bibfnamefont
  {L.~A.}\ \bibnamefont {Demianets}},\ }\href@noop {} {\bibfield  {journal}
  {\bibinfo  {journal} {Physica B}\ }\textbf {\bibinfo {volume} {354}},\
  \bibinfo {pages} {297} (\bibinfo {year} {2004})}\BibitemShut {NoStop}%
\bibitem [{\citenamefont {Svistov}\ \emph {et~al.}(2003)\citenamefont
  {Svistov}, \citenamefont {Smirnov}, \citenamefont {Prozorova}, \citenamefont
  {Petrenko}, \citenamefont {Demianets},\ and\ \citenamefont
  {Shapiro}}]{Svi03}%
  \BibitemOpen
  \bibfield  {author} {\bibinfo {author} {\bibfnamefont {L.~E.}\ \bibnamefont
  {Svistov}}, \bibinfo {author} {\bibfnamefont {A.~I.}\ \bibnamefont
  {Smirnov}}, \bibinfo {author} {\bibfnamefont {L.~A.}\ \bibnamefont
  {Prozorova}}, \bibinfo {author} {\bibfnamefont {O.~A.}\ \bibnamefont
  {Petrenko}}, \bibinfo {author} {\bibfnamefont {L.~N.}\ \bibnamefont
  {Demianets}}, \ and\ \bibinfo {author} {\bibfnamefont {A.~Y.}\ \bibnamefont
  {Shapiro}},\ }\href@noop {} {\bibfield  {journal} {\bibinfo  {journal} {Phys.
  Rev. B}\ }\textbf {\bibinfo {volume} {67}},\ \bibinfo {pages} {094434}
  (\bibinfo {year} {2003})}\BibitemShut {NoStop}%
\bibitem [{\citenamefont {Svistov}\ \emph {et~al.}(2006)\citenamefont
  {Svistov}, \citenamefont {Smirnov}, \citenamefont {Prozorova}, \citenamefont
  {Petrenko}, \citenamefont {Micheler}, \citenamefont {B\"uttgen},
  \citenamefont {Shapiro},\ and\ \citenamefont {Demianets}}]{Svi06}%
  \BibitemOpen
  \bibfield  {author} {\bibinfo {author} {\bibfnamefont {L.~E.}\ \bibnamefont
  {Svistov}}, \bibinfo {author} {\bibfnamefont {A.~I.}\ \bibnamefont
  {Smirnov}}, \bibinfo {author} {\bibfnamefont {L.~A.}\ \bibnamefont
  {Prozorova}}, \bibinfo {author} {\bibfnamefont {O.~A.}\ \bibnamefont
  {Petrenko}}, \bibinfo {author} {\bibfnamefont {A.}~\bibnamefont {Micheler}},
  \bibinfo {author} {\bibfnamefont {N.}~\bibnamefont {B\"uttgen}}, \bibinfo
  {author} {\bibfnamefont {A.~Y.}\ \bibnamefont {Shapiro}}, \ and\ \bibinfo
  {author} {\bibfnamefont {L.~N.}\ \bibnamefont {Demianets}},\ }\href@noop {}
  {\bibfield  {journal} {\bibinfo  {journal} {Phys. Rev. B}\ }\textbf {\bibinfo
  {volume} {74}},\ \bibinfo {pages} {024412} (\bibinfo {year}
  {2006})}\BibitemShut {NoStop}%
\bibitem [{\citenamefont {Smirnov}\ \emph {et~al.}(2007)\citenamefont
  {Smirnov}, \citenamefont {Yashiro}, \citenamefont {Kimura}, \citenamefont
  {Hagiwara}, \citenamefont {Narumi}, \citenamefont {Kindo}, \citenamefont
  {Kikkawa}, \citenamefont {Katsumata}, \citenamefont {Shapiro},\ and\
  \citenamefont {Demianets}}]{Smi07}%
  \BibitemOpen
  \bibfield  {author} {\bibinfo {author} {\bibfnamefont {A.~I.}\ \bibnamefont
  {Smirnov}}, \bibinfo {author} {\bibfnamefont {H.}~\bibnamefont {Yashiro}},
  \bibinfo {author} {\bibfnamefont {S.}~\bibnamefont {Kimura}}, \bibinfo
  {author} {\bibfnamefont {M.}~\bibnamefont {Hagiwara}}, \bibinfo {author}
  {\bibfnamefont {Y.}~\bibnamefont {Narumi}}, \bibinfo {author} {\bibfnamefont
  {K.}~\bibnamefont {Kindo}}, \bibinfo {author} {\bibfnamefont
  {A.}~\bibnamefont {Kikkawa}}, \bibinfo {author} {\bibfnamefont
  {K.}~\bibnamefont {Katsumata}}, \bibinfo {author} {\bibfnamefont {A.~Y.}\
  \bibnamefont {Shapiro}}, \ and\ \bibinfo {author} {\bibfnamefont {L.~N.}\
  \bibnamefont {Demianets}},\ }\href@noop {} {\bibfield  {journal} {\bibinfo
  {journal} {Phys. Rev. B}\ }\textbf {\bibinfo {volume} {75}},\ \bibinfo
  {pages} {134412} (\bibinfo {year} {2007})}\BibitemShut {NoStop}%
\bibitem [{\citenamefont {Kenzelmann}\ \emph {et~al.}(2007)\citenamefont
  {Kenzelmann}, \citenamefont {Lawes}, \citenamefont {Harris}, \citenamefont
  {Gasparovic}, \citenamefont {Broholm}, \citenamefont {Ramirez}, \citenamefont
  {Jorge}, \citenamefont {Jaime}, \citenamefont {Park}, \citenamefont {Huang},
  \citenamefont {Shapiro},\ and\ \citenamefont {Demianets}}]{Ken07}%
  \BibitemOpen
  \bibfield  {author} {\bibinfo {author} {\bibfnamefont {M.}~\bibnamefont
  {Kenzelmann}}, \bibinfo {author} {\bibfnamefont {G.}~\bibnamefont {Lawes}},
  \bibinfo {author} {\bibfnamefont {A.~B.}\ \bibnamefont {Harris}}, \bibinfo
  {author} {\bibfnamefont {G.}~\bibnamefont {Gasparovic}}, \bibinfo {author}
  {\bibfnamefont {C.}~\bibnamefont {Broholm}}, \bibinfo {author} {\bibfnamefont
  {A.~P.}\ \bibnamefont {Ramirez}}, \bibinfo {author} {\bibfnamefont {G.~A.}\
  \bibnamefont {Jorge}}, \bibinfo {author} {\bibfnamefont {M.}~\bibnamefont
  {Jaime}}, \bibinfo {author} {\bibfnamefont {S.}~\bibnamefont {Park}},
  \bibinfo {author} {\bibfnamefont {Q.}~\bibnamefont {Huang}}, \bibinfo
  {author} {\bibfnamefont {A.~Y.}\ \bibnamefont {Shapiro}}, \ and\ \bibinfo
  {author} {\bibfnamefont {L.~A.}\ \bibnamefont {Demianets}},\ }\href@noop {}
  {\bibfield  {journal} {\bibinfo  {journal} {Phys. Rev. Lett.}\ }\textbf
  {\bibinfo {volume} {98}},\ \bibinfo {pages} {267205} (\bibinfo {year}
  {2007})}\BibitemShut {NoStop}%
\bibitem [{\citenamefont {Hearmon}\ \emph {et~al.}(2012)\citenamefont
  {Hearmon}, \citenamefont {Fabrizi}, \citenamefont {Chapon}, \citenamefont
  {Johnson}, \citenamefont {Prabhakaran}, \citenamefont {Streltsov},
  \citenamefont {Brown},\ and\ \citenamefont {Radaelli}}]{Hea12}%
  \BibitemOpen
  \bibfield  {author} {\bibinfo {author} {\bibfnamefont {A.~J.}\ \bibnamefont
  {Hearmon}}, \bibinfo {author} {\bibfnamefont {F.}~\bibnamefont {Fabrizi}},
  \bibinfo {author} {\bibfnamefont {L.~C.}\ \bibnamefont {Chapon}}, \bibinfo
  {author} {\bibfnamefont {R.~D.}\ \bibnamefont {Johnson}}, \bibinfo {author}
  {\bibfnamefont {D.}~\bibnamefont {Prabhakaran}}, \bibinfo {author}
  {\bibfnamefont {S.~V.}\ \bibnamefont {Streltsov}}, \bibinfo {author}
  {\bibfnamefont {P.~J.}\ \bibnamefont {Brown}}, \ and\ \bibinfo {author}
  {\bibfnamefont {P.~G.}\ \bibnamefont {Radaelli}},\ }\href@noop {} {\bibfield
  {journal} {\bibinfo  {journal} {Phys. Rev. Lett.}\ }\textbf {\bibinfo
  {volume} {108}},\ \bibinfo {pages} {237201} (\bibinfo {year}
  {2012})}\BibitemShut {NoStop}%
\bibitem [{\citenamefont {Harris}(2007)}]{Har07}%
  \BibitemOpen
  \bibfield  {author} {\bibinfo {author} {\bibfnamefont {A.~B.}\ \bibnamefont
  {Harris}},\ }\href@noop {} {\bibfield  {journal} {\bibinfo  {journal} {Phys.
  Rev. B}\ }\textbf {\bibinfo {volume} {76}},\ \bibinfo {pages} {054447}
  (\bibinfo {year} {2007})}\BibitemShut {NoStop}%
\bibitem [{\citenamefont {Inami}(2007)}]{Ina07}%
  \BibitemOpen
  \bibfield  {author} {\bibinfo {author} {\bibfnamefont {T.}~\bibnamefont
  {Inami}},\ }\href@noop {} {\bibfield  {journal} {\bibinfo  {journal} {J.
  Solid State Chem.}\ }\textbf {\bibinfo {volume} {180}},\ \bibinfo {pages}
  {2075} (\bibinfo {year} {2007})}\BibitemShut {NoStop}%
\bibitem [{\citenamefont {Johnson}\ \emph {et~al.}(2011)\citenamefont
  {Johnson}, \citenamefont {Nair}, \citenamefont {Chapon}, \citenamefont
  {Bombardi}, \citenamefont {Vecchini}, \citenamefont {Prabhakaran},
  \citenamefont {Boothroyd},\ and\ \citenamefont {Radaelli}}]{Joh11}%
  \BibitemOpen
  \bibfield  {author} {\bibinfo {author} {\bibfnamefont {R.~D.}\ \bibnamefont
  {Johnson}}, \bibinfo {author} {\bibfnamefont {S.}~\bibnamefont {Nair}},
  \bibinfo {author} {\bibfnamefont {L.~C.}\ \bibnamefont {Chapon}}, \bibinfo
  {author} {\bibfnamefont {A.}~\bibnamefont {Bombardi}}, \bibinfo {author}
  {\bibfnamefont {C.}~\bibnamefont {Vecchini}}, \bibinfo {author}
  {\bibfnamefont {D.}~\bibnamefont {Prabhakaran}}, \bibinfo {author}
  {\bibfnamefont {A.~T.}\ \bibnamefont {Boothroyd}}, \ and\ \bibinfo {author}
  {\bibfnamefont {P.~G.}\ \bibnamefont {Radaelli}},\ }\href@noop {} {\bibfield
  {journal} {\bibinfo  {journal} {Phys. Rev. Lett.}\ }\textbf {\bibinfo
  {volume} {107}},\ \bibinfo {pages} {137205} (\bibinfo {year}
  {2011})}\BibitemShut {NoStop}%
\bibitem [{\citenamefont {Klevtsova}\ and\ \citenamefont
  {Klevtsov}(1970)}]{Kle70}%
  \BibitemOpen
  \bibfield  {author} {\bibinfo {author} {\bibfnamefont {R.~F.}\ \bibnamefont
  {Klevtsova}}\ and\ \bibinfo {author} {\bibfnamefont {P.~V.}\ \bibnamefont
  {Klevtsov}},\ }\href@noop {} {\bibfield  {journal} {\bibinfo  {journal}
  {Kristallografiya}\ }\textbf {\bibinfo {volume} {15}},\ \bibinfo {pages}
  {209} (\bibinfo {year} {1970})}\BibitemShut {NoStop}%
\bibitem [{Sup({\natexlab{a}})}]{Sup2}%
  \BibitemOpen
  \href@noop {} {} ({\natexlab{a}}),\ \bibinfo {note} {this `effective'
  inter-plane interaction $J_{\rm p}$ would by itself lead to commensurate
  interplane order.}\BibitemShut {Stop}%
\bibitem [{\citenamefont {Vajk}\ \emph {et~al.}(2005)\citenamefont {Vajk},
  \citenamefont {Kenzelmann}, \citenamefont {Lynn}, \citenamefont {Kim},\ and\
  \citenamefont {Cheong}}]{Vaj05}%
  \BibitemOpen
  \bibfield  {author} {\bibinfo {author} {\bibfnamefont {O.~P.}\ \bibnamefont
  {Vajk}}, \bibinfo {author} {\bibfnamefont {M.}~\bibnamefont {Kenzelmann}},
  \bibinfo {author} {\bibfnamefont {J.~W.}\ \bibnamefont {Lynn}}, \bibinfo
  {author} {\bibfnamefont {S.~B.}\ \bibnamefont {Kim}}, \ and\ \bibinfo
  {author} {\bibfnamefont {S.-W.}\ \bibnamefont {Cheong}},\ }\href@noop {}
  {\bibfield  {journal} {\bibinfo  {journal} {Phys. Rev. Lett.}\ }\textbf
  {\bibinfo {volume} {94}},\ \bibinfo {pages} {087601} (\bibinfo {year}
  {2005})}\BibitemShut {NoStop}%
\bibitem [{\citenamefont {Harris}\ \emph {et~al.}(1992)\citenamefont {Harris},
  \citenamefont {Kallin},\ and\ \citenamefont {Berlinsky}}]{Har92}%
  \BibitemOpen
  \bibfield  {author} {\bibinfo {author} {\bibfnamefont {A.~B.}\ \bibnamefont
  {Harris}}, \bibinfo {author} {\bibfnamefont {C.}~\bibnamefont {Kallin}}, \
  and\ \bibinfo {author} {\bibfnamefont {A.~J.}\ \bibnamefont {Berlinsky}},\
  }\href@noop {} {\bibfield  {journal} {\bibinfo  {journal} {Phys. Rev. B}\
  }\textbf {\bibinfo {volume} {45}},\ \bibinfo {pages} {2899} (\bibinfo {year}
  {1992})}\BibitemShut {NoStop}%
\bibitem [{Sup({\natexlab{b}})}]{Sup}%
  \BibitemOpen
  \href@noop {} {} ({\natexlab{b}}),\ \bibinfo {note} {see Supplemental
  Material section for more details.}\BibitemShut {Stop}%
\bibitem [{\citenamefont {Ribeiro}\ and\ \citenamefont
  {Perez-Mato}(2011)}]{Rib11}%
  \BibitemOpen
  \bibfield  {author} {\bibinfo {author} {\bibfnamefont {J.~L.}\ \bibnamefont
  {Ribeiro}}\ and\ \bibinfo {author} {\bibfnamefont {J.~M.}\ \bibnamefont
  {Perez-Mato}},\ }\href@noop {} {\bibfield  {journal} {\bibinfo  {journal} {J.
  Phys.: Condens. Matter}\ }\textbf {\bibinfo {volume} {23}},\ \bibinfo {pages}
  {446003} (\bibinfo {year} {2011})}\BibitemShut {NoStop}%
\bibitem [{\citenamefont {Penc}\ \emph {et~al.}(2004)\citenamefont {Penc},
  \citenamefont {Shannon},\ and\ \citenamefont {Shiba}}]{Pen04}%
  \BibitemOpen
  \bibfield  {author} {\bibinfo {author} {\bibfnamefont {K.}~\bibnamefont
  {Penc}}, \bibinfo {author} {\bibfnamefont {N.}~\bibnamefont {Shannon}}, \
  and\ \bibinfo {author} {\bibfnamefont {H.}~\bibnamefont {Shiba}},\
  }\href@noop {} {\bibfield  {journal} {\bibinfo  {journal} {Phys. Rev. Lett.}\
  }\textbf {\bibinfo {volume} {93}},\ \bibinfo {pages} {197203} (\bibinfo
  {year} {2004})}\BibitemShut {NoStop}%
\bibitem [{\citenamefont {Sergienko}\ and\ \citenamefont
  {Dagotto}(2006)}]{Ser06}%
  \BibitemOpen
  \bibfield  {author} {\bibinfo {author} {\bibfnamefont {I.~A.}\ \bibnamefont
  {Sergienko}}\ and\ \bibinfo {author} {\bibfnamefont {E.}~\bibnamefont
  {Dagotto}},\ }\href@noop {} {\bibfield  {journal} {\bibinfo  {journal} {Phys.
  Rev. B}\ }\textbf {\bibinfo {volume} {73}},\ \bibinfo {pages} {094434}
  (\bibinfo {year} {2006})}\BibitemShut {NoStop}%
\bibitem [{\citenamefont {Katsura}\ \emph {et~al.}(2005)\citenamefont
  {Katsura}, \citenamefont {Nagaosa},\ and\ \citenamefont {Balatsky}}]{Kat05}%
  \BibitemOpen
  \bibfield  {author} {\bibinfo {author} {\bibfnamefont {H.}~\bibnamefont
  {Katsura}}, \bibinfo {author} {\bibfnamefont {N.}~\bibnamefont {Nagaosa}}, \
  and\ \bibinfo {author} {\bibfnamefont {A.~V.}\ \bibnamefont {Balatsky}},\
  }\href@noop {} {\bibfield  {journal} {\bibinfo  {journal} {Phys. Rev. Lett.}\
  }\textbf {\bibinfo {volume} {95}},\ \bibinfo {pages} {057205} (\bibinfo
  {year} {2005})}\BibitemShut {NoStop}%
\bibitem [{\citenamefont {Kaplan}\ and\ \citenamefont {Mahanti}(2011)}]{Kap11}%
  \BibitemOpen
  \bibfield  {author} {\bibinfo {author} {\bibfnamefont {T.~A.}\ \bibnamefont
  {Kaplan}}\ and\ \bibinfo {author} {\bibfnamefont {S.~D.}\ \bibnamefont
  {Mahanti}},\ }\href@noop {} {\bibfield  {journal} {\bibinfo  {journal} {Phys.
  Rev. B}\ }\textbf {\bibinfo {volume} {83}},\ \bibinfo {pages} {174432}
  (\bibinfo {year} {2011})}\BibitemShut {NoStop}%
\end{thebibliography}%

\end{document}